\documentclass[12pt]{article}
  \usepackage[english]{babel}
\usepackage[pdftex]{color}
\usepackage{amsmath}
\usepackage{graphicx}
\usepackage{hyperref}
\usepackage{amsmath}
\usepackage{amsfonts}
\usepackage{amssymb}
\begin{document}

\title{
{\bf{} On Lagrangian formulations for arbitrary
\\bosonic HS fields on
Minkowski backgrounds}}

\author{\sc Alexander A.  Reshetnyak\thanks{reshet@ispms.tsc.ru}
  \\[0.5cm]
  Laboratory of Non-Linear Media Physics, Institute of
\\  Strength Physics and Materials Science, 634021 Tomsk, Russia}
\date{}

\maketitle 

\begin{abstract}
We review the details of  unconstrained Lagrangian formulations
for  Bose particles propagated on an arbitrary dimensional flat
space-time and described by the unitary irreducible integer
higher-spin representations of
 the Poincare
group subject to  Young tableaux $Y(s_1,...,s_k)$ with  $k$ rows.
The procedure is based on the construction of Verma modules and
finding auxiliary oscillator realizations for the symplectic
$sp(2k)$ algebra which encodes the second-class operator
constraints subsystem  in the HS symmetry algebra. Application of
an universal BRST approach reproduces gauge-invariant Lagrangians
with reducible gauge symmetries describing the free dynamics of
both massless and massive bosonic fields of any spin with
appropriate number of auxiliary fields.
\end{abstract}

\section{Introduction}

Growth of the interest to higher-spin (HS) field theory is mainly
stipulated  by the hopes to reconsider the problems of an unique
description of variety of elementary particles and all known
interactions especially due to expected output of LHC on the
planned capacity. Remind, that it suspects both the proof of
supersymmetry display, the answer on the question on existence of
Higgs boson, and probably a new insight on origin of Dark Matter
(\cite{LHC}). Because of HS field theory is closely related to
superstring theory, which operates with an infinite tower of
bosonic and fermionic HS fields it  can be treated as an approach
to study a structure of superstring theory from field-theoretic
viewpoint. Some of the aspects of current state of HS field theory
are discussed in the reviews \cite{reviews}. The paper considers
the last results of constructing
 Lagrangian formulations (LFs) for free integer  HS fields on flat
$\mathbb{R}^{1,d-1}$-space-time subject to arbitrary Young
tableaux $Y(s_1,...,s_k)$ in Fronsdal metric-like formalism within
BFV-BRST approach \cite{BFV}, and  based on the results presented
in \cite{BuchbinderReshetnyak} (see
Ref.\cite{BuchbinderReshetnyak}, for detailed bibliography on
various aspects).

It is known that for $d>4$ space-time dimensions, there appear,
besides totally symmetric irreducible representations of Poincare
or (Anti)-de-Sitter ((A)dS) algebras the mixed-symmetry
representations determined by more than one spin-like parameters
\cite{Labastida},  \cite{metsaevmixirrep}. Whereas  for the former
ones the LFs both for massless and massive free higher-spin fields
is well enough developed \cite{massless Minkowski}, \cite{massless
AdS1}, \cite{massless AdS}, \cite{massive Minkowski},
\cite{massive AdS} for the latter  the problem of their
field-\-theoretic description is not completely solved. So,
 the main result within the problem  of unconstrained LF for
 arbitrary massless mixed-symmetry HS fields on a Minkowski space-time
 was obtained in \cite{SkvortsovLF} with use of unfolded form of
 equations of motion for the field in "frame-like"
 formulation. In the "metric-like" formulation corresponding
 Lagrangians were  derived in closed manner for only
 reducible Poincare group $ISO(1,d-1)$ representations in
 \cite{Franciamix}. In turn,
the Labastida-like \cite{Labastida} constrained LF with off-shell
traceless and mixed-symmetry holonomic constraints for arbitrary
mixed-symmetric higher spin fields  were recently studied in
\cite{KostyaMaxim} and on a basis of detour complexes from the
BRST quantization of worldline diffeomorphism invariant systems in
\cite{WaldronBRST}.

The paper is devoted to the construction of unconstrained general
gauge-invariant Lagrangians for both massless and massive
mixed-symmetry tensor fields of rank $s_1 + s_2 + ... + s_k$, with
any integer numbers $s_1 \geq s_2 \geq ... \geq s_k \geq 1$ for $k
\leq [d/2]$ in a $\mathbb{R}^{1,d-1}$ space, the fields being
elements of Poincare-group  irreps with a Young tableaux (YT)
having $k$ rows. Our approach is based on the BFV--BRST
construction \cite{BFV} which
 in application to
free HS field theory consists of three steps:  transforming of
conditions that determine the representations with a given spin
into a topological  gauge system of mixed-class operator
constraints in an auxiliary Fock space. Then, the subsystem of the
second-class constraints, is converted, with a preservation of the
initial algebraic structure, into a system of first-class
constraints alone in an enlarged Fock space  with respect to which
one constructs the BRST charge. Finally, the Lagrangian for a HS
field is constructed in terms of the BRST charge in such a way
that, first, the corresponding equations of motion reproduce the
initial constraints, second, the LF  contain all appropriate
auxiliary and Stuckelberg fields. The BRST approach to LF of HS
field theories has been developed for arbitrary massless and
massive  bosonic  fields in Minkowski and AdS spaces in
\cite{0001195}--\cite{symferm-ads}.

\section{Integer
HS Symmetry Algebra  for Bosonic fields}\label{Symmalgebra}   A
massless integer spin irreducible representation of Poincare group
in  $\mathbb{R}^{1,d-1}$  is described by a tensor field
$\Phi_{(\mu^1)_{s_1},(\mu^2)_{s_2},...,(\mu^k)_{s_k}}
\hspace{-0.2em}\equiv \hspace{-0.2em}
\Phi_{\mu^1_1\ldots\mu^1_{s_1},\mu^2_1\ldots\mu^2_{s_2},...,}$
${}_{ \mu^k_1\ldots \mu^k_{s_k}}(x)$
 of rank $\sum_{i\geq 1}^k s_i$ and generalized spin
 $\mathbf{s} = (s_1, s_2,
 ... , s_k)$, ($s_1 \geq s_2\geq ... \geq s_k>0, k
\leq [d/2])$ subject to a YT with $k$ rows of respective length
$s_1, s_2, ..., s_k$
\begin{equation}\label{Young k}
\Phi_{(\mu^1)_{s_1},(\mu^2)_{s_2},...,(\mu^k)_{s_k}}
\hspace{-0.3em}\longleftrightarrow \hspace{-0.3em}
\begin{array}{|c|c|c c c|c|c|c|c|c| c|}\hline
  \!\mu^1_1 \!&\! \mu^1_2\! & \cdot \ & \cdot \ & \cdot \ & \cdot\  & \cdot\  & \cdot\ &
  \cdot\    &\!\! \mu^1_{s_1}\!\! \\
   \hline
    \! \mu^2_1\! &\! \mu^2_2\! & \cdot\
   & \cdot\ & \cdot  & \cdot &  \cdot & \!\!\mu^2_{s_2}\!\!   \\
  \cline{1-8} \!\!\cdots\!\!   \\
   \cline{1-7}
    \! \mu^k_1\! &\! \mu^k_2\! & \cdot\
   & \cdot\ & \cdot  & \cdot &   \!\!\mu^k_{s_k}\!\!   \\
   \cline{1-7}
\end{array}\ ,
\end{equation}
This field is symmetric with respect to the permutations of each
type of Lorentz indices $\mu^i$, (for the signature $\eta_{\mu\nu}
= diag (+, -,...,-)$, $\mu, \nu = 0,1,...,d-1$)
  and
obeys to the Klein-Gordon,  divergentless (\ref{Eq-1b}), traceless
(\ref{Eq-2b}) and mixed-symmetry equations (\ref{Eq-3b}) [for
$i,j=1,...,k;\, l_i,m_i=1,...,s_i$]:
\begin{eqnarray}
&&
\partial^\mu\partial_\mu\Phi_{(\mu^1)_{s_1},(\mu^2)_{s_2},...,(\mu^k)_{s_k}}
 =0, \qquad
\partial^{\mu^i_{l_i}}\Phi_{
(\mu^1)_{s_1},(\mu^2)_{s_2},...,(\mu^k)_{s_k}} =0,  \label{Eq-1b}
\\ && \eta^{\mu^i_{l_i}\mu^i_{m_i}}\Phi_{
(\mu^1)_{s_1},(\mu^2)_{s_2},...,(\mu^k)_{s_k}}=
\eta^{\mu^i_{l_i}\mu^j_{m_j}}\Phi_{
(\mu^1)_{s_1},(\mu^2)_{s_2},...,(\mu^k)_{s_k}} =0, \quad
 l_i<m_i,  \label{Eq-2b}
 \end{eqnarray}
\vspace{-4ex}
\begin{eqnarray}
&& \Phi_{
(\mu^1)_{s_1},...,\{(\mu^i)_{s_i}\underbrace{,...,\mu^j_{1}...}\mu^j_{l_j}\}...\mu^j_{s_j},...(\mu^k)_{s_k}}=0,\quad
i<j,\ 1\leq l_j\leq s_j, \label{Eq-3b}
\end{eqnarray}
where the  bracket below denote that the indices  in it do not
include in  symmetrization, i.e. the symmetrization concerns only
indices $(\mu^i)_{s_i}, \mu^j_{l_j} $ in
$\{(\mu^i)_{s_i}\underbrace{,...,\mu^j_{1}...}\mu^j_{l_j}\}$.

Simultaneous description of all  $ISO(1,d-1)$ group irreps maybe
reformulated in a standard manner with an auxiliary Fock space
$\mathcal{H}$, generated by $k$ pairs of bosonic creation
$a^i_{\mu^i}(x)$ and annihilation $a^{j+}_{\nu^j}(x)$ operators,
$i,j =1,...,k, \mu^i,\nu^j =0,1...,d-1$: $[a^i_{\mu^i},
a_{\nu^j}^{j+}]=-\eta_{\mu^i\nu^j}\delta^{ij}$
 and a set of constraints for an arbitrary string-like (so called \emph{basic}) vector
$|\Phi\rangle \in \mathcal{H}$,
\begin{eqnarray}
\label{PhysState}  \hspace{-2ex}&& \hspace{-2ex} |\Phi\rangle  =
\sum_{s_1=0}^{\infty}\sum_{s_2=0}^{s_1}\cdots\sum_{s_k=0}^{s_{k-1}}
\Phi_{(\mu^1)_{s_1},(\mu^2)_{s_2},...,(\mu^k)_{s_k}}(x)\,
\prod_{i=1}^k\prod_{l_i=1}^{s_i} a^{+\mu^i_{l_i}}_i|0\rangle,\\
\label{l0} \hspace{-2ex}&&  \bigl(l_0, {l}^i, l^{ij}, t^{i_1j_1}
\bigr)|\Phi\rangle = \bigl(\partial^\mu\partial_\mu, -i a^i_\mu
\partial^\mu, \textstyle\frac{1}{2}a^{i}_\mu a^{j\mu},
a^{i_1+}_\mu a^{j_1\mu}\bigr) |\Phi\rangle=0,\ i\leq j;\, i_1 <
j_1.
\end{eqnarray}
 The set of $(k(k+1)+1)$
primary constraints (\ref{l0}), $\{o_\alpha\}$ = $\bigl\{{{l}}_0,
{l}^i, l^{ij}, t^{i_1j_1} \bigr\}$, are equivalent to Eqs.
(\ref{Eq-1b})--(\ref{Eq-3b}) for all spins. In turn, additional
condition, $g_0^i|\Phi\rangle =(s_i+\frac{d}{2}) |\Phi\rangle$ for
 number particles operators, $g_0^i = - a^{i+}_\mu  a^{\mu{}i} +
 \frac{d}{2}$, makes (\ref{l0}) to
be equivalent to Eqs. (\ref{Eq-1b})--(\ref{Eq-3b}) for given spin
$\mathbf{s} = (s_1, s_2,  ... , s_k)$.

The procedure of LF  implies  the Hermiticity  of BFV-BRST
operator $Q$, $Q = C^\alpha o_\alpha + \ldots$,   that means the
extension of the set $\{o_\alpha\}$ up to one of $\{o_I\} =
\{o_\alpha, o^+_\alpha; g_0^i\}$, which is closed with respect to
hermitian conjugation related to standard scalar product on
$\mathcal{H}$ and commutator multiplication $[\ ,\ ]$. Operators
$o_I$ satisfy to the Lie-algebra commutation relations,
    $[o_I,\ o_J]= f^K_{IJ}o_K$, for structure constants
    $f^K_{IJ}= - f^K_{JI}$, to be determined from the
multiplication table~\ref{table in}. \hspace{-1ex}{\begin{table}
{{\footnotesize
\begin{center}
\begin{tabular}{||c||c|c|c|c|c|c|c||c||}\hline\hline
$\hspace{-0.2em}[\; \downarrow, \rightarrow
]\hspace{-0.5em}$\hspace{-0.7em}&
 $t^{i_1j_1}$ & $t^+_{i_1j_1}$ &
$l_0$ & $l^i$ &$l^{i{}+}$ & $l^{i_1j_1}$ &$l^{i_1j_1{}+}$ &
$g^i_0$ \\
\hline\hline $t^{i_2j_2}$
    & $A^{i_2j_2, i_1j_1}$ & $B^{i_2j_2}{}_{i_1j_1}$
   & $0$&\hspace{-0.3em}
    $\hspace{-0.2em}l^{j_2}\delta^{i_2i}$\hspace{-0.5em} &
    \hspace{-0.3em}
    $-l^{i_2+}\delta^{j_2 i}$\hspace{-0.3em}
    &\hspace{-0.7em} $\hspace{-0.7em}l^{\{j_1j_2}\delta^{i_1\}i_2}
    \hspace{-0.9em}$ \hspace{-1.2em}& \hspace{-1.2em}$
    -l^{i_2\{i_1+}\delta^{j_1\}j_2}\hspace{-0.9em}$\hspace{-1.2em}& $F^{i_2j_2,i}$ \\
\hline $t^+_{i_2j_2}$
    & $-B^{i_1j_1}{}_{i_2j_2}$ & $A^+_{i_1j_1, i_2j_2}$
&$0$   & \hspace{-0.3em}
    $\hspace{-0.2em} l_{i_2}\delta^{i}_{j_2}$\hspace{-0.5em} &
    \hspace{-0.3em}
    $-l^+_{j_2}\delta^{i}_{i_2}$\hspace{-0.3em}
    & $l_{i_2}{}^{\{j_1}\delta^{i_1\}}_{j_2}$ & $-l_{j_2}{}^{\{j_1+}
    \delta^{i_1\}}_{i_2}$ & $-F_{i_2j_2}{}^{i+}$\\
\hline $l_0$
    & $0$ & $0$
& $0$   &
    $0$\hspace{-0.5em} & \hspace{-0.3em}
    $0$\hspace{-0.3em}
    & $0$ & $0$ & $0$ \\
\hline $l^j$
   & \hspace{-0.5em}$- l^{j_1}\delta^{i_1j}$ \hspace{-0.5em} &
   \hspace{-0.5em}$
   -l_{i_1}\delta_{j_1}^{j}$ \hspace{-0.9em}  & \hspace{-0.3em}$0$ \hspace{-0.3em} & $0$&
   \hspace{-0.3em}
   $l_0\delta^{ji}$\hspace{-0.3em}
    & $0$ & \hspace{-0.5em}$- \textstyle\frac{1}{2}l^{\{i_1+}\delta^{j_1\}j}$
    \hspace{-0.9em}&$l^j\delta^{ij}$  \\
\hline $l^{j+}$ & \hspace{-0.5em}$l^{i_1+}
   \delta^{j_1j}$\hspace{-0.7em} & \hspace{-0.7em}
   $l_{j_1}^+\delta_{i_1}^{j}$ \hspace{-1.0em} &
   $0$&\hspace{-0.3em}
      \hspace{-0.3em}
   $-l_0\delta^{ji}$\hspace{-0.3em}
    \hspace{-0.3em}
   &\hspace{-0.5em} $0$\hspace{-0.5em}
    &\hspace{-0.7em} $ \textstyle\frac{1}{2}l^{\{i_1}\delta^{j_1\}j}
    $\hspace{-0.7em} & $0$ &$-l^{j+}\delta^{ij}$  \\
\hline $l^{i_2j_2}$
    & \hspace{-0.3em}$\hspace{-0.4em}-l^{j_1\{j_2}\delta^{i_2\}i_1}\hspace{-0.5em}$
    \hspace{-0.5em} &\hspace{-0.5em} $\hspace{-0.4em}
    -l_{i_1}{}^{\{i_2+}\delta^{j_2\}}_{j_1}\hspace{-0.3em}$\hspace{-0.3em}
   & $0$&\hspace{-0.3em}
    $0$\hspace{-0.5em} & \hspace{-0.3em}
    $ \hspace{-0.7em}-\textstyle\frac{1}{2}l^{\{i_2}\delta^{j_2\}i}
    \hspace{-0.5em}$\hspace{-0.3em}
    & $0$ & \hspace{-0.7em}$\hspace{-0.3em}
L^{i_2j_2,i_1j_1}\hspace{-0.3em}$\hspace{-0.7em}& $\hspace{-0.7em}  l^{i\{i_2}\delta^{j_2\}i}\hspace{-0.7em}$\hspace{-0.7em} \\
\hline $l^{i_2j_2+}$
    & $ l^{i_1 \{i_2+}\delta^{j_2\}j_1}$ & $ l_{j_1}{}^{\{j_2+}
    \delta^{i_2\}}_{i_1}$
   & $0$&\hspace{-0.3em}
    $\hspace{-0.2em} \textstyle\frac{1}{2}l^{\{i_2+}\delta^{ij_2\}}$\hspace{-0.5em} & \hspace{-0.3em}
    $0$\hspace{-0.3em}
    & $-L^{i_1j_1,i_2j_2}$ & $0$ &$\hspace{-0.5em}  -l^{i\{i_2+}\delta^{j_2\}i}\hspace{-0.3em}$\hspace{-0.2em} \\
\hline\hline $g^j_0$
    & $-F^{i_1j_1,j}$ & $F_{i_1j_1}{}^{j+}$
   &$0$& \hspace{-0.3em}
    $\hspace{-0.2em}-l^i\delta^{ij}$\hspace{-0.5em} & \hspace{-0.3em}
    $l^{i+}\delta^{ij}$\hspace{-0.3em}
    & \hspace{-0.7em}$\hspace{-0.7em}  -l^{j\{i_1}\delta^{j_1\}j}\hspace{-0.7em}$\hspace{-0.7em} & $ l^{j\{i_1+}\delta^{j_1\}j}$&$0$ \\
   \hline\hline
\end{tabular}
\end{center}}} \vspace{-2ex}\caption{HS symmetry  algebra  $\mathcal{A}(Y(k),
\mathbb{R}^{1,d-1})$.\label{table in} }\end{table}

The products $B^{i_2j_2}_{i_1j_1}, A^{i_2j_2, i_1j_1},
F^{i_1j_1,i}, L^{i_2j_2,i_1j_1}$ in the table~\ref{table in} are
given by the  relations,
\begin{eqnarray}
  &&\hspace{-3em}{}B^{i_2j_2}{}_{i_1j_1}\ =\
  (g_0^{i_2}-g_0^{j_2})\delta^{i_2}_{i_1}\delta^{j_2}_{j_1} +
  (t_{j_1}{}^{j_2}\theta^{j_2}{}_{j_1} + t^{j_2}{}^+_{j_1}\theta_{
  j_1}{}^{j_2})\delta^{i_2}_{i_1}
  -(t^+_{i_1}{}^{i_2}\theta^{i_2}{}_{i_1} + t^{i_2}{}_{i_1}\theta_{i_1}{}^{
  i_2})
  \delta^{j_2}_{j_1}
\,,\label{Bijkl}
\\
   &&\hspace{-3em}A^{i_2j_2, i_1j_1} =  t^{i_1j_2}\delta^{i_2j_1}-
  t^{i_2j_1}\delta^{i_1j_2}  ,\qquad  F^{i_2j_2,i} \ = \
   t^{i_2j_2}(\delta^{j_2i}-\delta^{i_2i}),\label{Fijk} \\
  &&\hspace{-3em}L^{i_2j_2,i_1j_1} =   \textstyle\frac{1}{4}\hspace{-0.15em}
  \Bigl\{\delta^{i_2i_1}
\delta^{j_2j_1}\hspace{-0.15em}\Bigl[\hspace{-0.15em}
2g_0^{i_2}\delta^{i_2j_2} \hspace{-0.15em}+\hspace{-0.15em}
g_0^{i_2}\hspace{-0.15em} +\hspace{-0.15em}
g_0^{j_2}\hspace{-0.15em}\Bigr] \hspace{-0.15em} -
\hspace{-0.15em}\Bigl(\hspace{-0.15em}\delta^{j_2\{i_1}\hspace{-0.15em}
\Bigl[\hspace{-0.15em}t^{j_1\}i_2}\theta^{i_2j_1\}}
\hspace{-0.15em}+\hspace{-0.15em}t^{i_2j_1\}+}\theta^{j_1\}i_2}\hspace{-0.15em}\Bigr]\hspace{-0.15em}
\hspace{-0.15em}+ \hspace{-0.15em}(j_2\leftrightarrow
i_2)\hspace{-0.15em}\Bigr)\hspace{-0.25em}\Bigr\} , \label{Lklij}
\end{eqnarray}
with Heaviside $\theta$-symbol $\theta^{ij}$. From the Hamiltonian
analysis of the dynamical systems
 the operators $\{o_I\}$ contain respective $2k^2$ second-class $\{o_a\}=\{l^{ij},t^{i_1j_1}, l^+_{ij},t^+_{i_1j_1}
 \}$,
   $(2k+1)$ first-class $\{l_0, l^{i}, l^+_{j} \}$ constraints
  subsystems and $k$ elements $g_0^i$ which forms the
  non-vanishing in $\mathcal{H}$ matrix $\Delta_{ab}(g_0^i)$ in $[o_a,o_b] \sim
  \Delta_{ab}$  for topological gauge system.

We called in \cite{BuchbinderReshetnyak} the algebra of the
operators $O_I$ as \emph{integer higher-spin symmetry algebra in
Minkowski space with a Young tableaux having $k$ rows} and denoted
it as $\mathcal{A}(Y(k), \mathbb{R}^{1,d-1})$.

The subsystem of the second-class constraints $\{o_a\}$ together
with $\{g_0^i\}$ forms the subalgebra in $\mathcal{A}(Y(k),
\mathbb{R}^{1,d-1})$ which is isomorphic to the symplectic
$sp(2k)$ algebra in turn to be Howe dual \cite{Howe1} to the
Lorentz algebra $so(1,d-1)$ (see details in
\cite{BuchbinderReshetnyak} and in \cite{KostyaMaxim} as well).
Note, the elements $g_0^i$, form a basis in the Cartan subalgebra
whereas $l^{ij}, {t}_{i}{}^j$ are the basis of low-triangular
subalgebra in $sp(2k)$.

Having constructed the HS symmetry algebra, we  can  not still
construct BRST operator $Q$ with respect to the elements $o_I$
from  $\mathcal{A}(Y(k), \mathbb{R}^{1,d-1})$ due to second-class
constraints $\{o_a\}$ presence in it. One should to convert
symplectic algebra $sp(2k)$ of $\{o_a,g_0^i\}$  into enlarged set
of operators $O_I$ with only first-class constraints.

\section{Verma module and Oscillator realization for $sp(2k)$
}\label{Vermamodule}

We consider an additive conversion procedure developed within BRST
approach, (see e.g. \cite{BurdikPashnev}),  which implies the
enlarging of $o_I$ to $O_I = o_I + o'_I$, with additional parts
$o'_I$ to be given on a new Fock space $\mathcal{H}'$ being
independent on $\mathcal{H}'$. In this case the elements $O_I$ are
given on $\mathcal{H}\otimes \mathcal{H}'$ so that the requirement
for $O_I$ to be in involution, i.e. $[O_I,\ O_J] \sim O_K$,  leads
to the series of the same algebraic relations, for $O_I$ and
$o'_I$ as those for $o_I$.

Leaving aside the details of Verma module (special representation
space \cite{Dixmier}) construction for the symplectic algebra
$sp(2k)$ of new operators $o'_I$ considered in
\cite{BuchbinderReshetnyak}, we  present here theirs explicit
oscillator form in terms of new $2k^2$ creation and annihilation
operators $(B^c;B^+_d)$ = $(b^{+}_{ij}, d^+_{rs}; b_{ij},
d_{rs})$, $i,j,r,s =1,\ldots, k; i\leq j; r<s$ as follows,
\begin{eqnarray}
g_0^{\prime i}& = &  \sum_{l\leq m}
 b_{lm}^+b_{lm}(\delta^{il}+\delta^{im}) + \sum_{r< s}d^+_{rs}d_{rs}(\delta^{is}-
 \delta^{ir}) +h^i
 \,,\label{g'0iF} \\
l^{\prime+}_{ij} & = & b_{ij}^+\,\qquad  t^{\prime+}_{lm}   =
d^+_{lm} - \sum_{n=1}^{l-1}d_{nl}d^+_{nm}
   - \sum_{n=1}^{k}(1+\delta_{nl})b^+_{nm}b_{ln}\,,
 \label{t'+lmtext}
 \\
t^{\prime }_{lm} &=& - \sum_{n=1}^{l-1}d^+_{nl}d_{nm} +
\sum_{p=0}^{m-l-1}\sum_{k_1=l+1}^{m-1}\ldots \sum_{k_p=l+p}^{m-1}
 C^{k_{p}m}(d^+,d)\prod_{j=1}^pd_{k_{j-1}k_{j}}
 \\
  && -\sum_{n=1}^{k}(1+\delta_{nm})b^+_{nl}
b_{nm}
 \label{t'lmF}\,, \qquad k_0\equiv l,\texttt{ where }\nonumber\\
  \label{Clm}
C^{lm}(d^+,d)&\equiv &
\Bigl(h^{l}-h^{m}-\sum_{n=m}^{k}\bigl(d^+_{ln}d_{ln}+d^+_{mn}d_{mn}\bigl)+
\sum_{n=l+1}^{m-1}d^+_{nm}d_{nm}-d^+_{lm}d_{lm}\Bigr)d_{lm} \\
 &&  - \sum_{n=l+1}^{m-1}d^+_{ln}d_{nm} + \sum_{n=m+1}^{k}\Bigl\{d^+_{mn}  - \sum_{n'=1}^{m-1} d^+_{n'n} d_{n'm}\Bigr\}d_{ln},\texttt{ for }l<m.\nonumber
\end{eqnarray}
Note, first, that $B_c, B^+_d$ satisfy to the standard (only
nonvanishing) commutation relations, $[B_c, B^+_d]= \delta_{cd}$,
second, the arbitrary parameters $h^i$ in (\ref{g'0iF}) serve to
reproduce correct LF fo HS field with given spin $\mathbf{s}$,
whereas the form of the rest elements $l^{\prime }_{ij}$, for
$i\leq j$, to be expressed by means of $C^{lm}(d^+,d)$ may be
found in \cite{BuchbinderReshetnyak}.  At last, the operators
$l^{\prime+}_{ij}, t^{\prime+}_{lm}$ are not Hermitian conjugated
respectively $l^{\prime}_{ij}, t^{\prime}_{lm}$ in $\mathcal{H}'$
with respect to standard scalar product as in $\mathcal{H}$. To
restore that property we introduce new scalar product,
\begin{align}
& \langle{\Phi}_1|K'E^{- \prime\alpha}|\Phi_2\rangle =
\langle{\Phi}_2|K'E^{\prime\alpha}|\Phi_1\rangle^* ,  &&
\langle{\Phi}_1|K'g_0^{\prime i}|\Phi_2\rangle =
\langle{\Phi}_2|K'g_0^{\prime i}|\Phi_1\rangle^*,
\end{align}
for   $(E^{\prime\alpha};E^{-\prime\alpha}) = (l^{\prime }_{ij},
t^{\prime }_{lm}; l^{\prime +}_{ij}, t^{\prime +}_{lm})$ with
 the operator $K'$, relating standard scalar products in Verma
 module and  in $\mathcal{H}'$ \cite{BuchbinderReshetnyak}\footnote{The
 case of the massive  bosonic HS fields whose
system of second-class constraints contains additionally to
elements of $sp(2k)$ algebra  the constraints  of isometry
subalgebra of  Minkowski space  $l^i, l^+_i, l_0$ may be treated
via procedure of dimensional reduction of the algebra
$\mathcal{A}(Y(k),\mathbb{R}^{1,d})$ for massless HS fields to one
$\mathcal{A}(Y(k),\mathbb{R}^{1,d-1})$ for massive HS fields, (see
\cite{BuchbinderReshetnyak}). Now, the wave equation in
(\ref{Eq-1b}) is changed on Klein-Gordon equation corresponding to
the constraint $l_0$ ($l_0=\partial^\mu\partial_\mu +m^2$)  acting
on the same  \emph{basic} vector $|\Phi\rangle$
(\ref{PhysState}).}.

\section{BRST-BFV operator and Lagrangian formulations}
\label{BRSToperator}
Because of the algebra of $O_I$ under consideration is a Lie
algebra $\mathcal{A}(Y(k),\mathbb{R}^{1,d-1})$ the BFV-BRST
operator $Q'$ can be constructed in the standard way by the
formula,
\begin{equation}\label{generalQ'}
    Q'  = {O}_I\mathcal{C}^I + \frac{1}{2}
    \mathcal{C}^I\mathcal{C}^Jf^K_{JI}\mathcal{P}_K
\end{equation}
with the constants $f^K_{JI}$ from the table~\ref{table in}, for
the constraints $O_I = (L_0, L^+_i$, $L_i$, $L_{ij}, L^+_{ij},
T_{ij}$, $T^+_{ij}$, $G_0^i)$, fermionic  ghost fields and
conjugated to them momenta $(C^I, \mathcal{P}_I)$  =
$\bigl((\eta_0, {\cal{}P}_0); (\eta^i, {\cal{}P}^+_i)$;
$(\eta^+_i, {\cal{}P}_j); (\eta^{ij}, {\cal{}P}^+_{ij})$;
$(\eta^+_{ij}, {\cal{}P}_{ij}); (\vartheta_{rs},\lambda^+_{rs})$;
$(\vartheta^+_{rs}, \lambda_{rs}); (\eta^i_{G},
{\cal{}P}_{G})\bigr)$  with the properties
\begin{equation}\label{propgho}
    \eta^{ij}= \eta^{ji} ,  \vartheta_{rs}= \vartheta_{rs}\theta^{sr}
,\ \{\vartheta_{rs},\lambda^+_{tu}\}= \delta_{rt}\delta_{su},\
\{{\cal{}P}_j, \eta_i^+\}=\delta_{ij},\
\{\eta_{lm},{\cal{}P}_{ij}^+\}= \delta_{li}\delta_{jm}
\end{equation}
 and non-vanishing anticommutators $\{\eta_0,{\cal{}P}_0\}= \imath,\
\{\eta^i_{\mathcal{G}}, {\cal{}P}^j_{\mathcal{G}}\}
 = \imath\delta^{ij}$ for zero-mode ghosts\footnote{The ghosts possess the
 standard  ghost number distribution,
$gh(\mathcal{C}^I)$ = $ - gh(\mathcal{P}_I)$ = $1$
$\Longrightarrow$  $gh({Q}')$ = $1$.}. The property of the BRST
operator to be Hermitian is defined by the rule with respect to
the scalar product $\langle \ |\ \rangle$ in $\mathcal{H}_{tot} =
\mathcal{H} \otimes \mathcal{H}' \otimes \mathcal{H}_{gh}$,
\begin{eqnarray}\label{HermQ}
  Q^{\prime +}K = K Q'\,,\texttt{ for }K =  \hat{1} \otimes K' \otimes
  \hat{1}_{gh}.
  \end{eqnarray}
To construct LF for bosonic  HS fields in a $\mathbb{R}^{1,d-1}$
Minkowski  space we partially follow the algorithm of
\cite{BurdikPashnev}, \cite{BuchKrycRysTak}, which is a particular
case of our construction, corresponding to $s_3 = 0$. First, we
extract the dependence of  $Q'$ (\ref{generalQ'}) on the ghosts
$\eta^i_{G}, {\cal{}P}^i_{G}$, to obtain the BRST operator $Q$
only for the system of converted first-class constraints $\{O_I\}
\setminus \{G^i_0\}$ and generalized spin operator $\sigma^i$:
\begin{eqnarray}
\label{Q'} {Q}' \hspace{-0.4em} &=& Q +
\eta^i_{G}(\sigma^i+h^i)+\mathcal{A}^i \mathcal{P}^i_{G},\quad
\texttt{ where }\\
\label{Q} {Q} \hspace{-0.4em} &=&\hspace{-0.4em} \textstyle
 \frac{1}{2}\eta_0L_0+\eta_i^+L^i
+\sum\limits_{l\leq m}\eta_{lm}^+L^{lm} + \sum\limits_{l<
m}\vartheta^+_{lm}T^{lm}
 + \frac{\imath}{2}\sum_l\eta_l^+\eta^l{\cal{}P}_0-
\sum\limits_{l<n<m}\vartheta_{lm}^+\vartheta^{l}{}_n\lambda^{nm}
 \nonumber
\\\hspace{-0.4em}
&& {}\hspace{-1.5em}+
\sum\limits_{n<l<m}\vartheta_{lm}^+\vartheta_{n}{}^m\lambda^{+nl}
-\sum\limits_{i<l<j} (\vartheta^+_{lj}\vartheta^+_{i}{}^l -
\vartheta^+_{il}\vartheta^{+l}{}_{j})\lambda^{ij} -
\sum_{n,l<m}(1+\delta_{ln})\vartheta_{lm}^+\eta^{l+}{}_{n}
\mathcal{P}^{mn}
\nonumber\\
\hspace{-0.4em}&& \hspace{-1.5em}+
\sum_{n,l<m}(1+\delta_{mn})\vartheta_{lm}^+\eta^{m}{}_{n}
\mathcal{P}^{+ln}+ \textstyle\frac{1}{2}\sum\limits_{l<m,n\leq
m}\eta^+_{nm}\eta^{n}{}_l\lambda^{lm}-
\bigl[\textstyle\frac{1}{2}\sum\limits_{l\leq
m}(1+\delta_{lm})\eta^m\eta_{lm}^+
\nonumber\\
\hspace{-0.4em} && \hspace{-1.2em}  
 +
\sum\limits_{l<m}\vartheta_{lm} \eta^{+m}
+\sum\limits_{m<l}\vartheta^+_{ml} \eta^{+m} \bigr]\mathcal{P}^l
+h.c. \quad\texttt{ and} \\
\label{sigmai}
  \sigma^i &=& G_0^i - h^i   - \eta_i \mathcal{P}^+_i +
   \eta_i^+ \mathcal{P}_i + \sum_{
m}(1+\delta_{im})(
\eta_{im}^+{\cal{}P}^{im}-\eta_{im}{\cal{}P}^+_{im})\nonumber\\
   &&  + \sum_{l<i}[\vartheta^+_{li}
\lambda^{li} - \vartheta^{li}\lambda^+_{li}]-
\sum_{i<l}[\vartheta^+_{il} \lambda^{il} -
\vartheta^{il}\lambda^+_{il}]\,,
\end{eqnarray}
with some inessential for LF operatorial quantities
$\mathcal{A}^i$. Then, we choose a representation of
$\mathcal{H}_{tot}$ as, $(\eta_i, \eta_{ij},  \vartheta_{rs},
\mathcal{P}_0, \mathcal{P}_i, \mathcal{P}_{ij}, \lambda_{rs},
\mathcal{P}^{i}_G)|0\rangle=0$ and suppose that the field vectors
$|\chi \rangle$ as well as the gauge parameters $|\Lambda \rangle$
do not depend on ghosts $\eta^{i}_G$
\begin{eqnarray}
|\chi \rangle &=& \sum_n \prod_{i\le j, r<s}^k( b_{ij}^+
)^{n_{ij}}( d_{rs}^+ )^{p_{rs}}( \eta_0^+ )^{n_{f 0}}\prod_{i, j,
l\le m, n\le o}( \eta_i^+ )^{n_{f i}} ( \mathcal{P}_j^+ )^{n_{p
j}} ( \eta_{lm}^+ )^{n_{f lm}} ( \mathcal{P}_{no}^+ )^{n_{pno}}
\nonumber
\\
&&{} \times\prod\nolimits_{r<s, t<u}( \vartheta_{rs}^+)^{n_{f rs}}
( \lambda_{tu}^+ )^{n_{\lambda tu}} |\Phi(a^+_i)^{n_{f 0} (n)_{f
i}(n)_{p j}(n)_{f lm} (n)_{pno}(n)_{f rs}(n)_{\lambda
tu}}_{(n)_{l}(n)_{ij}(p)_{rs}}\rangle\footnotemark \,. \label{chi}
\end{eqnarray}\footnotetext{The brackets $(n)_{f i},(n)_{p j}, (n)_{ij}$ in (\ref{chi}) means,
e.g., for $(n)_{ij}$ the set of indices $(n_{11},...$,
$n_{1k},..., n_{k1},..., n_{kk})$. The  sum above is taken over
$n_{l}$, $n_{ij}$, $p_{rs}$ and  running from $0$ to infinity, and
over the rest $n$'s from $0$ to $1$.}We denote by $|\chi^k\rangle$
the state (\ref{chi}) satisfying to $gh(|\chi^k\rangle)=-k$. Thus,
the physical state having the ghost number zero is
$|\chi^0\rangle$, the gauge parameters $|\Lambda \rangle$ having
the ghost number $-1$ is $|\chi^1\rangle$ and so on. The vector
$|\chi^0\rangle$ must contain  physical string-like vector
$|\Phi\rangle = |\Phi(a^+_i)^{(0)_{f o} (0)_{f i}(0)_{p j}(0)_{f
lm} (0)_{pno}(0)_{f rs}(0)_{\lambda tu}}_{
(0)_{ij}(0)_{rs}}\rangle$:
\begin{eqnarray}\label{decomptot}
|\chi^0\rangle&=&|\Phi\rangle+  |\Phi_A\rangle ,\texttt{ where }
|\Phi_A\rangle_{\big|[B^+_a = C^I = \mathcal{P}_I = 0]} =0
\end{eqnarray}
Independence of the vectors (\ref{chi}) on $\eta^{i}_G$ transforms
the equation for the physical state ${Q}'|\chi^0\rangle=0$ and the
BRST complex of the reducible gauge transformations,
$\delta|\chi\rangle$ = $Q'|\chi^1\rangle$, $\delta|\chi^1\rangle =
Q'|\chi^2\rangle$, $\ldots$, $\delta|\chi^{(r-1)}\rangle =
Q'|\chi^{(r)}\rangle$, to the relations:
\begin{align}
\label{Qchi} & 1)\ Q|\chi^0\rangle=0, && 2)\
\delta|\chi^0\rangle=Q|\chi^1\rangle, &&\ldots\quad
r)\ \delta|\chi^{r-1}\rangle=Q|\chi^{(r)}\rangle,\\
& 1)\ (\sigma^i+h^i)|\chi^0\rangle=0,
 &&2)\
(\sigma^i+h^i)|\chi^1\rangle=0, && \ldots\quad r)\
(\sigma^i+h^i)|\chi^{r}\rangle=0, \label{QLambda}
\end{align}
Where $r-1=k(k+1)-1$ is the stage of reducibility both for
massless and for the massive bosonic HS field. Resolving the
spectral  problem from the Eqs.(\ref{QLambda})
 we determine  the eigenvectors of
the operators $\sigma^i$: $|\chi^0\rangle_{(n)_k}$,
$|\chi^1\rangle_{(n)_k}$, $\ldots$, $|\chi^{s}\rangle_{(n)_k}$,
$n_1 \geq n_2 \geq \ldots n_k \geq 0$ and corresponding
eigenvalues of the parameters $h^i$ (for massless  HS fields),
\begin{eqnarray}
\label{hi} -h^i &=& n_i+\textstyle\frac{d-2-4i}{2} \;, \quad
i=1,..,k\,,\quad n_1,...,n_{k-1} \in \mathbb{Z}, n_k \in
\mathbb{N}_0\,,
\end{eqnarray}
Let us fix some values of $n_i=s_i$. Then one should substitute
$h^i$ corresponding to the chosen $s_i$ (\ref{hi})  into $K$
(\ref{HermQ}), $Q$ (\ref{Q}) and relations (\ref{Qchi}),
(\ref{QLambda}). Thus, e.g., the equation of motion in
(\ref{Qchi}) corresponding to the field with given spin
$(s_1,...,s_k)$ has the form, $Q_{(n)_k}|\chi^0\rangle_{(n)_k}=0$,
with nilpotent $Q_{(n)_k}$ and the same for the gauge
transformations  in (\ref{Qchi}).

Following to bosonic  one- \cite{0001195}, \cite{0505092} and
two-row cases , \cite{BurdikPashnev}, \cite{BuchKrycRysTak} one
can show that last equation may be derived from the Lagrangian
action for fixed spin $(n)_k=(s)_k$,
\begin{eqnarray}
\mathcal{S}_{(s)_k} = \int d \eta_0 \; {}_{(s)_k}\langle \chi^0
|K_{(s)_k} Q_{(s)_k}| \chi^0 \rangle_{(s)_k}, \
\Bigl(\Longrightarrow
 \frac{\delta \mathcal{S}_{(s)_k}}{\delta {}_{(s)_k}\langle \chi^0
|} =  Q_{(s)_k}| \chi^0 \rangle_{(s)_k} = 0\Bigr),\label{Scl}
\end{eqnarray}
where the standard scalar product for the creation and
annihilation operators in $\mathcal{H}_{tot}$ is assumed.

Concluding, one can prove  the Lagrangian action (\ref{Scl})
indeed reproduces the basic conditions
(\ref{Eq-1b})--(\ref{Eq-3b}) for massless  (massive) HS fields.
General action (\ref{Scl}) gives, in principle, a straight recept
to obtain the Lagrangian for any component field from general
vector $| \chi^0 \rangle_{(s)_k}$.

\paragraph{Acknowledgements} The author is thankful to the organizers
of the Dubna International Workshop SQS'11 for the hospitality.
A.R. is grateful to K. Alkalaev, M. Grigoriev, D. Francia, V.
Gershun, E. Latini, P.M. Lavrov, Yu.M. Zinoviev for valuable
discussions and comments and to I.L. Buchbinder for collaboration.
The work was supported by
 the RFBR grant, project  Nr. 12-02-00121
and by LRSS grant Nr.224.2012.2.

\end{document}